\documentclass{emulateapj}

\shorttitle{AMiBA Performance}
\shortauthors{Lin et al.}

\usepackage{graphicx}

\begin{document}

\title{AMiBA: System Performance}

\author{
Kai-Yang Lin\altaffilmark{1,2}, 
Chao-Te Li\altaffilmark{1}, 
Paul T.P. Ho\altaffilmark{1,3}, 
Chih-Wei Locutus Huang\altaffilmark{2,4}, 
Yu-Wei Liao\altaffilmark{2,4}, 
Guo-Chin Liu\altaffilmark{1,5}, 
Patrick M. Koch\altaffilmark{1}, 
Sandor M. Molnar\altaffilmark{1}, 
Hiroaki Nishioka\altaffilmark{1}, 
Keiichi Umetsu\altaffilmark{1,4}, 
Fu-Cheng Wang\altaffilmark{2,4}, 
Jiun-Huei Proty Wu\altaffilmark{2,4}, 
Michael Kestevan\altaffilmark{6}, 
Mark Birkinshaw\altaffilmark{7}, 
Pablo Altamirano\altaffilmark{1}, 
Chia-Hao Chang\altaffilmark{1}, 
Shu-Hao Chang\altaffilmark{1}, 
Su-Wei Chang\altaffilmark{1}, 
Ming-Tang Chen\altaffilmark{1}, 
Pierre Martin-Cocher\altaffilmark{1}, 
Chih-Chiang Han\altaffilmark{1}, 
Yau-De Huang\altaffilmark{1}, 
Yuh-Jing Hwang\altaffilmark{1}, 
Fabiola Iba\~nez-Roman\altaffilmark{1}, 
Homin Jiang\altaffilmark{1}, 
Derek Y. Kubo\altaffilmark{1}, 
Peter Oshiro\altaffilmark{1}, 
Philippe Raffin\altaffilmark{1}, 
Tashun Wei\altaffilmark{1},
Warwick Wilson\altaffilmark{6}, 
Ke-Jung Chen\altaffilmark{1}, 
and Tzihong Chiueh\altaffilmark{2,4}
}

\altaffiltext{1}{Academia Sinica Institute of Astronomy and Astrophysics, P.O. Box 23-141, Taipei, Taiwan 106}
\altaffiltext{2}{Physics Department, National Taiwan University, Taipei, Taiwan 106}
\altaffiltext{3}{Harvard-Smithsonian Center for Astrophysics, 60 Garden Street, Cambridge, MA 02138, USA}
\altaffiltext{4}{Leung Center for Cosmology and Particle Astrophysics, National Taiwan University, Taipei, Taiwan 106}
\altaffiltext{5}{Department of Physics, Tamkang University, 251-37 Tamsui, Taipei County, Taiwan}
\altaffiltext{6}{Australia Telescope National Facility, Epping, NSW Australia 1710}
\altaffiltext{7}{University of Bristol, Tyndall Avenue, Bristol BS8 1TL, UK}

\begin{abstract}
 The Y.T. Lee Array for Microwave Background Anisotropy (AMiBA) started
 scientific operation in early 2007. This work describes the optimization of 
 the system performance for the measurements of the Sunyaev-Zel'dovich effect 
 for six massive galaxy clusters at redshifts  $0.09 - 0.32$. 
 We achieved a point source sensitivity of $63\pm 7$ mJy
 with the seven 0.6~m dishes in 1 hour of on-source integration
 in 2-patch differencing observations.  
 We measured and compensated for the delays between the antennas of our
 platform-mounted interferometer. Beam switching was used to cancel
 instrumental instabilities and ground pick up.
 Total power and phase stability were good
 on time scales of hours, and the system was shown to integrate down on
 equivalent timescales of 300 hours per baseline/correlation, or about 10 hours
 for the entire array.  While the broadband correlator leads to
 good sensitivity, the small number of lags in the correlator resulted in
 poorly measured bandpass response.  We corrected for this by using external
 calibrators (Jupiter and Saturn). Using Jupiter as the flux standard, we 
 measured the disk brightness temperature of Saturn to be $149^{+5}_{-12}$~K.
\end{abstract}


\section{Introduction}
The angular power spectrum of cosmic microwave background (CMB)
anisotropies carries a wealth of information on the physical processes
in early epochs of the universe. A comparison of theoretical models
with accurate measurements of CMB anisotropies thus constrains the
fundamental cosmological parameters and models for cosmic structure
formation. On larger angular scales, the temperature anisotropies are
dominated by primary CMB fluctuations, whereas on smaller angular
scales secondary effects such as the Sunyaev-Zel'dovich (SZ) effects
due to galaxy clusters dominate over primordial anisotropies. The
amplitude and location of the peak in the thermal SZ power spectrum
are particularly sensitive to the amplitude of the primordial matter
power spectrum, represented by the normalization $\sigma_8$, as well
as the thermal history of the hot intracluster medium. 
The Cosmic Background Imager \citep[CBI,][]{pea03} and 
Arcminute Cosmology Bolometer Array Receiver \citep[ACBAR,][]{kuo04} 
measured the CMB temperature power spectrum at large angular multipoles of 
$l\sim3000$. 
While  the CBI detected an excess power over the theoretical prediction from 
the standard cosmological model, the ACBAR result has a larger error bar 
and is consistent with both an excess and no excess. 
To date the uncertainties of the high-$l$ measurements remain large. 
More accurate measurements on large
angular scales around and beyond $l = 3000$ are required 
to better constrain the value of $\sigma_8$
\citep[e.g.,][]{bon05,gol03,kyl04}. 

The Y.T. Lee Array for Microwave Background Anisotropy \citep[AMiBA,][]{ho08,
mtc08,koc08a} is designed to measure CMB anisotropies on these multipole 
scales. The AMiBA is located on the volcanic mountain Mauna Loa, Hawaii, at 
an altitude of 3400m. The array observes with a single sideband in
$86-102$~GHz, or at roughly 3~mm wavelength, with cooled HEMT low noise 
amplifiers (LNA). Each  
of the seven receivers measures two linear polarizations (X and Y) and 
produces two corresponding IF channels (each $2-18$~GHz). Out of the
four possible  
cross-correlations with a pair of receivers, AMiBA employs a switching system 
to form either the (XX$^{\ast}$, YY$^{\ast}$) or the (XY$^{\ast}$, YX$^{\ast}$)
product at the same time. Note that a circular polarizer is being developed so 
that AMiBA can choose to measure either the (LL$^{\ast}$, RR$^{\ast}$) or the 
(LR$^{\ast}$, RL$^{\ast}$) cross-correlations in the future. There are thus 
21 baselines and 42 instantaneous correlations for the seven-element array. 
The correlation is further divided into complex visibilities in two frequency 
bands using an analog four-lag correlator \citep{ctl04}. 

All antennas and receivers are mounted on a 6~m platform so that antennas can 
be closely packed without issues with shadowing and collision. 
In the 2007 and 2008 seasons, observations were made with 60~cm diameter 
dishes close-packed in the center of the platform. 
\citet{wu08} present details of the observations and analysis of six massive 
clusters. 

In this paper we describe how the system performance was optimized for these 
targeted observations. 
Two companion papers discuss the data integrity \citep{hn08} and the CMB and 
foreground uncertainty in the SZ flux estimation \citep{liu08}.  
Combined with published X-ray parameters, the SZ fluxes of six clusters were 
used to measure the Hubble parameter \citep{koc08b} and to examine the scaling 
relations \citep{hcw08}. Subaru weak lensing data for four of the
clusters were  
analyzed with the SZ measurements to derive the baryon fraction \citep{ku08}.

This paper is organized as follows. Critical issues such
as the noise temperatures, delay corrections, stability, spurious
signal removal and characteristics of the correlators are described in
\S\ref{SEC:OPTIMIZE}. \S\ref{SEC:PERFORMANCE} discusses the losses
of the system, the calibration errors, and the integration of
noise. Finally \S\ref{SEC:CONCLUSION} summarizes our conclusions.

\section{OPTIMIZING INTERFEROMETER PERFORMANCE\label{SEC:OPTIMIZE}}
Prior to and during the 2007 observing season, commissioning activities 
identified parts of the operations which needed to be improved \citep{kyl08}. 
In particular, \citet{ydh08} reports on the deformation of the platform which 
can affect the performance of the interferometer. Fortunately, these platform 
errors are repeatable and can be modeled. Their effects on pointing, radio 
alignment, and phase errors are discussed in \citet{koc08a}.  For AMiBA
operations in 2007-8 these effects were minimal.  In this paper, we concentrate
on other areas of the interferometer performance which were optimized.

\subsection{System Temperature}
To understand the gain stability of AMiBA, we first measured the receiver 
stabilities. The system temperature is monitored by a set of sky-dips in total 
power mode. The total power output from each IF channel can be approximated by 
\begin{eqnarray}
  P_{IF} &=& gkB [ T_{rx} + T_{dish} + T_{cmb} \nonumber\\
   & & \mbox{} + T_{atm}/\sin(el) + T_{gnd}(az,el) ],
\end{eqnarray}
where $g$ is the power gain, $k$ is the Boltzmann constant, $B$ is the
 bandwidth of each IF channel, and the $T$'s denote the noise
temperatures from the receiver ($rx$), antenna ($dish$), CMB ($cmb$),
the atmosphere ($atm$), and ground pickup ($gnd$). 
A hot/cold load measurement is used to calibrate $gB$ and $T_{rx}$.
The receiver noise temperatures are $55-75$~K \citep{mtc08}.
Fitting the total power to $P=P_0+P_1/\sin(el)$ lumps
the contributions into sky-like ($P_1$) and receiver-like ($P_0$)
parts plus some residual contributions from the ground.  
The measurements show that the total receiver-like noise temperature is
about $1 \sigma \sim 5$~K higher than $T_{rx}$.
The sky-like part is approximately $15$~K at zenith in typical
observing conditions. Including $T_{cmb}$, the system temperatures away 
from zenith are about $80-100$~K. 
Repeated hot/cold load measurements of the receiver noise temperatures show 
that $T_{rx}$ is stable within the measurement error ($\sim 5$~K).
Hence, by monitoring the system temperature using sky-dips, we can reject 
inferior sky conditions and unstable instrument behavior.

\subsection{Delay Correction\label{sec:delay}}
Since AMiBA is a coplanar array there is no fringe rotation in a
tracking observation. Fringes occur when a source moves across the
field of view (fov) creating a geometric delay. The fov of AMiBA
equipped with 0.6-m dishes is $23$\arcmin\ \citep{wu08}. The requirement on 
delay trimming is that the source delay should remain within the sampling
range of the lag-correlator, which is $\pm50$~ps. As the source delay
approaches the limit of sampling range, the error in the recovered
visibility becomes larger with a consequent rapid drop in
sensitivity. To allow a 2-m baseline to observe a $23$\arcmin~fov,
which corresponds to a delay range of $\sim\pm22$~ps, the instrumental
delay was specified to a tolerance of $\pm20$~ps. 

To measure the delay for each correlation, all dishes were removed and a noise 
source was mounted between receivers (e.g. Ant$_1$ and Ant$_2$). 
A fringe is generated when the noise source moves from Ant$_1$ toward Ant$_2$, 
simulating a fringe due to a celestial source.
\begin{equation}
  L(x,\tau_a) = \mathcal{R}\left( \int_{IF} df \mathbb{R^\prime}(f) e^{-i2\pi[(f+f_{LO})\frac{2x}{c} + f(\tau_2-\tau_1+\tau_a)]} \right)\,,
  \label{eq:tstage}
\end{equation}
where $x$ is the displacement of noise source, $f$ is the IF
frequency, and $\mathbb{R^\prime}$ is the complex response function of
the baseline excluding the linear part of the phase due to lags
($\tau_a$, $a=1...4$) in the correlator. $\mathcal{R}$ takes the real
part of the expression and is done implicitly whenever necessary
hereafter. $\tau_1$ and $\tau_2$ represent the instrumental delays in
the IF's of Ant$_1$ and Ant$_2$. The fringe envelope peaks when
$\frac{2x}{c}=\tau_1-\tau_2-\tau_a$. The relative delay
$\tau_1-\tau_2$ is measured with respect to the central lag (with
$\tau_a=0$). Equation (\ref{eq:tstage}) is usually referred to as the
lag output or the lag data throughout this work. 

We found the instrumental delays for all IF channels using relative
delay measurements. Short cables were then inserted into each IF for
compensation. After this trimming procedure the residual delays were
measured by fitting fringes for the Sun without the dishes,
modeling the fringes as the convolution of the observed point
source fringe with a circular disk. The differences between
observation and model are consistent with residual delays of $\pm
15$~ps (RMS).  Except for the delays due to platform deformation, the delays 
between antennas were therefore well controlled.

\subsection{Bandpass Shape Measurement}
\begin{figure*}
  \plotone{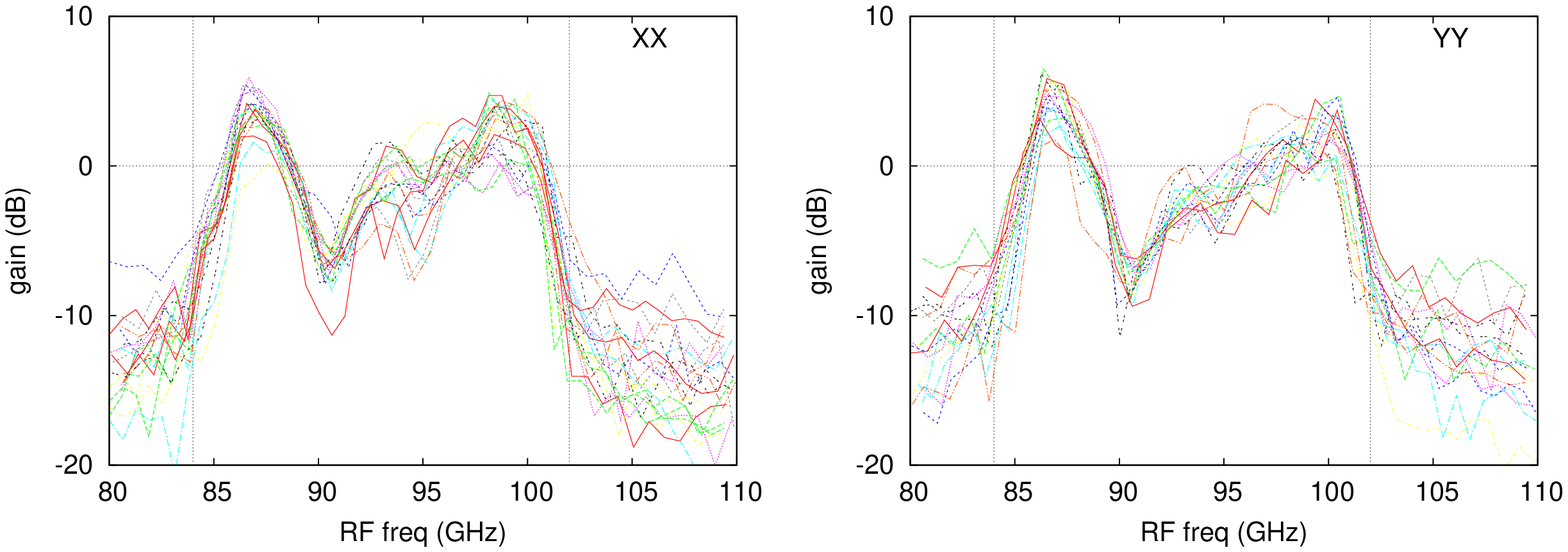}\\
  \plotone{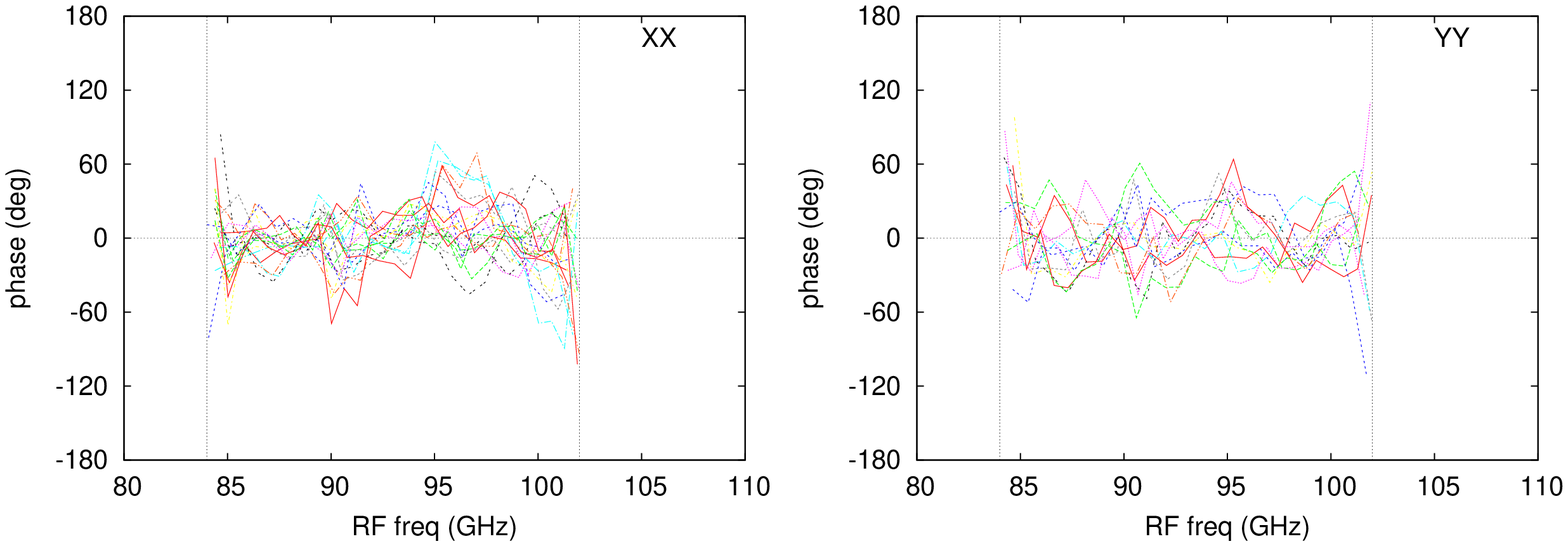}
  \caption{The complex responses of AMiBA. Responses include effects from the 
    RF components, IF components, and the analog correlator. Top and bottom 
    panels display the gain and phase responses respectively. Each line 
    represents one cross-correlation (XX on the left and YY on the right) of 
    a pair of receivers. Vertical dashed lines indicate RF frequencies
    of 84 and 102~GHz (IF frequencies 0 and 18~GHz).
  }
  \label{fig:responses}
\end{figure*}

The AMiBA correlator has four lags and outputs two spectral bands to
cover
the $2-18$~GHz band. Knowing the bandpass shape is an important aspect of 
obtaining good visibilities using this type of correlator (see next
section 
\S\ref{sec:lag2vis}). Because the analog correlator contributes significantly
to the bandpass shape, we adopted a baseline-based measurement
approach.
The Fourier transform of the fringe $L(x,\tau_a)$ against $x$ is
used to determine $\mathbb{R}^\prime$ for each baseline, with a
spectral sampling of about $0.8$~GHz. Each lag output is transformed 
independently. Fig.~\ref{fig:responses} displays the gain and phase responses 
of all valid measurements after the four lag outputs are averaged together. 
Averaging the phase responses of the four lags, the delays are canceled leaving
only the common mode of the spectral variation. The gain 
responses of the four lags are summed and then normalized such that the 
averaged gain between $2-18$~GHz is set to 1.  

The conversion from observed fringe rate to the RF frequency is proportional 
to the noise source translation speed. We believe a $\pm 1$~\% jitter is 
present in the translation stage we used, which introduced roughly $\pm 1$~GHz 
uncertainty in the response frequency. This causes one of the major problem 
in developing an accurate visibility extraction method from our lag
data (see \S\ref{sec:lag2vis}). An improved measurement setup involving 
simultaneous injection of a single frequency source is being developed to 
achieve a higher accuracy.

\begin{figure}
  \plotone{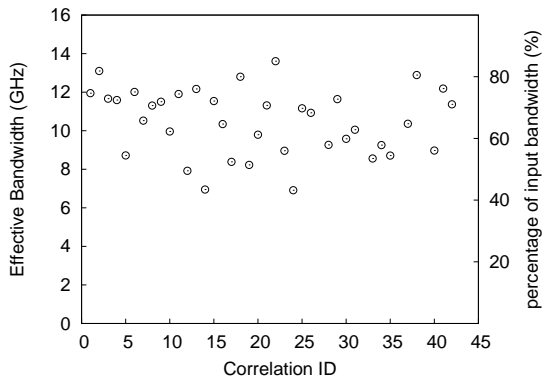}
  \caption{Effective bandwidths of the AMiBA correlators calculated from the 
   bandpasses displayed in Fig.\ref{fig:responses}. The percentages are
   based on a nominal input bandwidth of 16~GHz.
  }
  \label{fig:effbw}
\end{figure}

The effective bandwidths, defined as 
$B=|\int df \mathbb{R^\prime}|^2 / \int df |\mathbb{R^\prime}|^2$, 
are insensitive to the uncertainty in the response frequency. Based on
our bandpass measurements, the effective bandwidths of the AMiBA
correlators are calculated and shown in Fig.\ref{fig:effbw}. They
generally fall in the range of $7-13$~GHz.

\subsection{Extracting the Interferometer Visibilities \label{sec:lag2vis}}
Several approaches can be used to convert the four measured lags of the 
AMiBA correlator into
complex visibilities in two bands over the 16~GHz bandwidth. 
We find that the inaccuracies inherent in this inversion need to be corrected 
by external calibration.  Here we adopt the formalism of \citet{wu08}
\citep[see][for an alternative formalism]{ctl04}. The lag output in
equation~(\ref{eq:tstage}) can be expressed in matrix form as $L_a =
\mathbb{R}_{ak} \mathbb{V}^{src}_k$, where $\mathbb{V}^{src}_k$ is the
source visibility.  
Subscript $a$ indexes the $N_{lag}=4$ lags, and subscript $k$ indexes
the $N_{f}$ discretized frequency samples $f_k$, where $N_k$ is usually
much larger than $N_{lag}$.  

The transformation relies on a kernel $\mathbb{K}_{ak}$,
which is an estimate of the response matrix $\mathbb{R}_{ak}$. The
kernel is integrated in frequency into two bands
$\overline{\mathbb{K}}_{ac}$, where $c=1...4$ indexes the real and
imaginary parts of the two bands. We use the inverse of the
integrated kernel to construct the raw visibility 
$\mathbb{V}^{raw}_c \equiv {\overline{\mathbb{K}}^{-1}}_{ca} L_a$.

\begin{figure*}
  \plotone{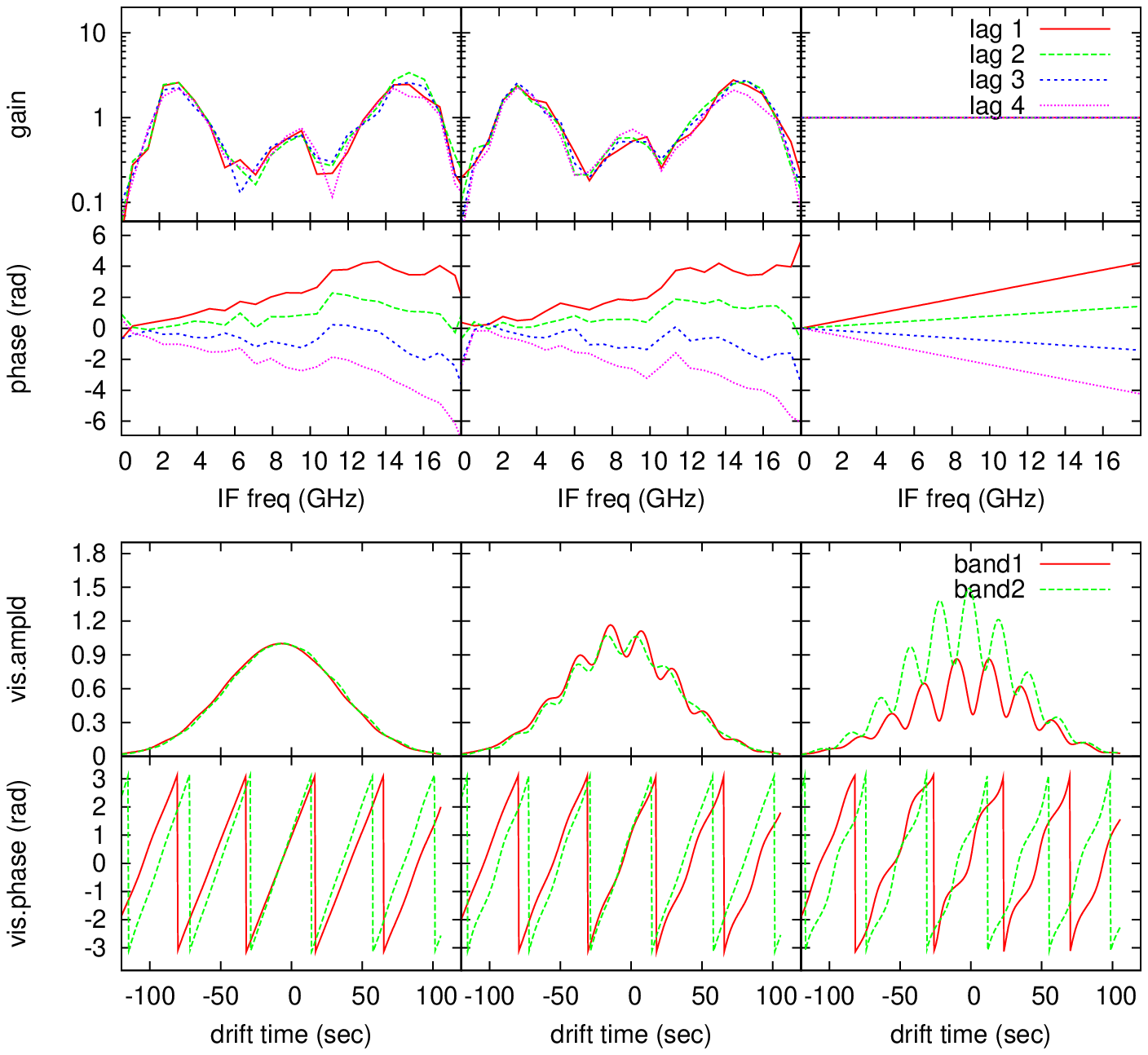}
  \caption{The upper two rows display three complex response functions.
    From the left, two slightly different functions are taken from the 
    measurements shown in Fig. \ref{fig:responses} with all four lags 
    plotted explicitly. The third is an assumed flat response function.
    We use the left-most function to simulate a set of fringes from a flat 
    spectrum source moving along the baseline. 
    Then we follow the lag-to-visibility procedures in \S\ref{sec:lag2vis}
    to obtain the bottom three sets of complex visibilities using 
    each of the response functions as the transformation kernel.
    The horizontal
    axis is the source offset presented as drift time. To the left we
    see the case when the kernel is the same as the response
    function. The central plots show the results with small errors in
    the kernel. On the right we show the result of using an assumed flat
    kernel. 
    Even though the deviation from ideal behavior can be as large as 
    50\% in visibilities, calibration can correct for the errors. 
  }
  \label{fig:lag2vis}
\end{figure*}

Ideally we would like $\mathbb{K}_{ak} = \mathbb{R}_{ak}$ so that 
$\mathbb{V}^{raw}_c$ is closest to $\mathbb{V}^{src}_c$.
However, when $\mathbb{K}_{ak}$ is an inaccurate
representation of $\mathbb{R}_{ak}$, due to measurement errors,
variations with temperature or time, insufficient spectral resolution
in the measurement, or insufficient information about the
response, errors in visibilities occur. 
Fig.~\ref{fig:lag2vis} demonstrates the calculation of raw
visibility using simulated drift scans in three cases when (1) the
kernel is the exact response, (2) there are measurement errors, and
(3) there is no knowledge about the response. The correct visibility
should appear as a Gaussian in amplitude with a linearly
increasing phase. It can be seen that case (1) recovers the correct
result, whereas deviations from this form increase with decreasing
accuracy of the kernel. We therefore must obtain a calibrated
visibility from the raw visibility
$\mathbb{V}^{cal}_{b} \equiv C_{bc} \mathbb{V}^{raw}_c$,
where $b$ has the same index range as $c$, and $C_{bc}$ is the
calibration matrix, which can be obtained by comparing the raw
visibility of a planet (the calibrator) to the theoretical visibility.

In the analysis of data taken in 2007 and 2008, the flat kernel (right-most
function in Fig.~\ref{fig:lag2vis}) was assumed, and planet
calibrations were applied. We have estimated the errors introduced by external 
calibration by running simulations on point source models.  This error is on 
the order of $\pm 2$\% (1$\sigma$) in the absolute fluxes with no detectable 
bias. This is small compared to the thermal noise and the measurement errors 
on the planet itself.

\subsection{Stability\label{sec:stability}}
\begin{figure*}
  \centering
  \includegraphics[height=3.6in]{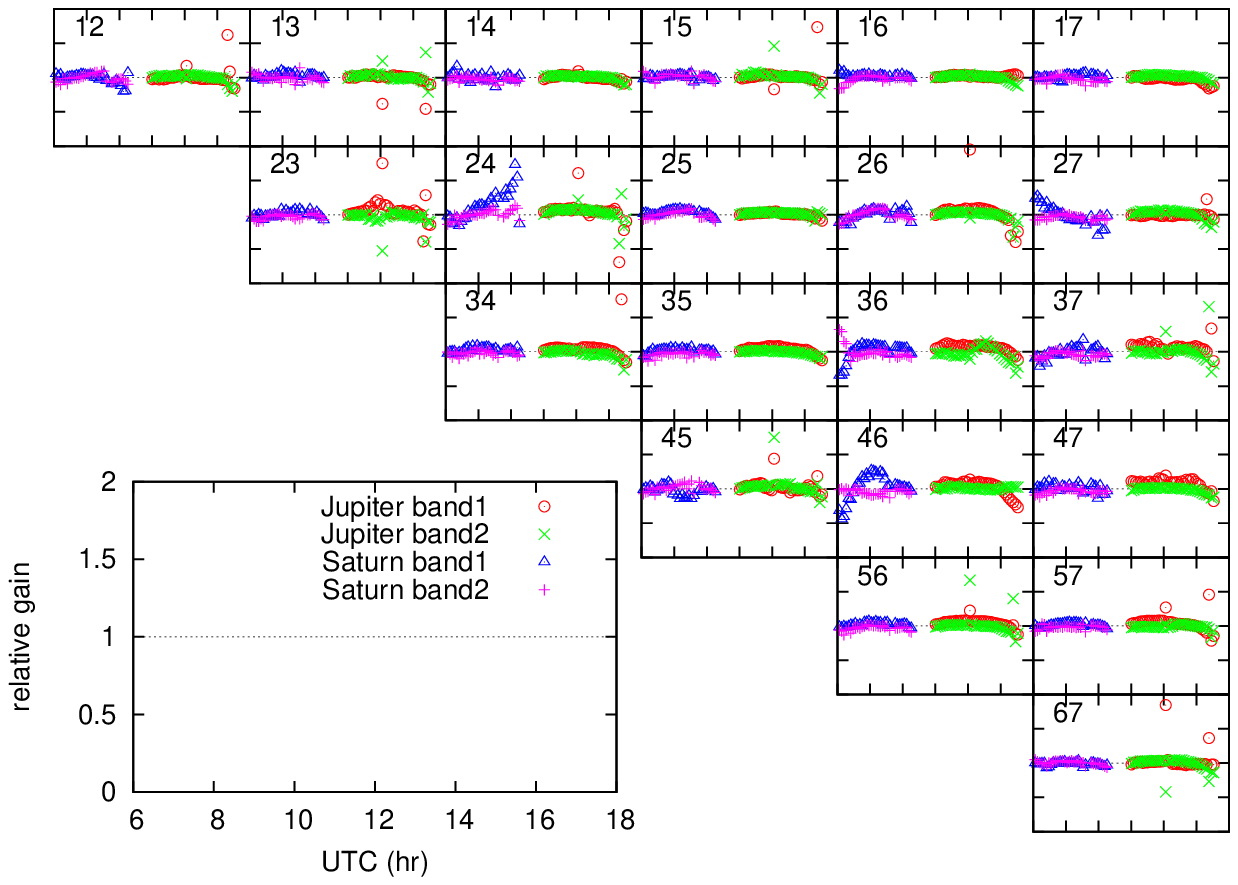}\\
  \includegraphics[height=3.6in]{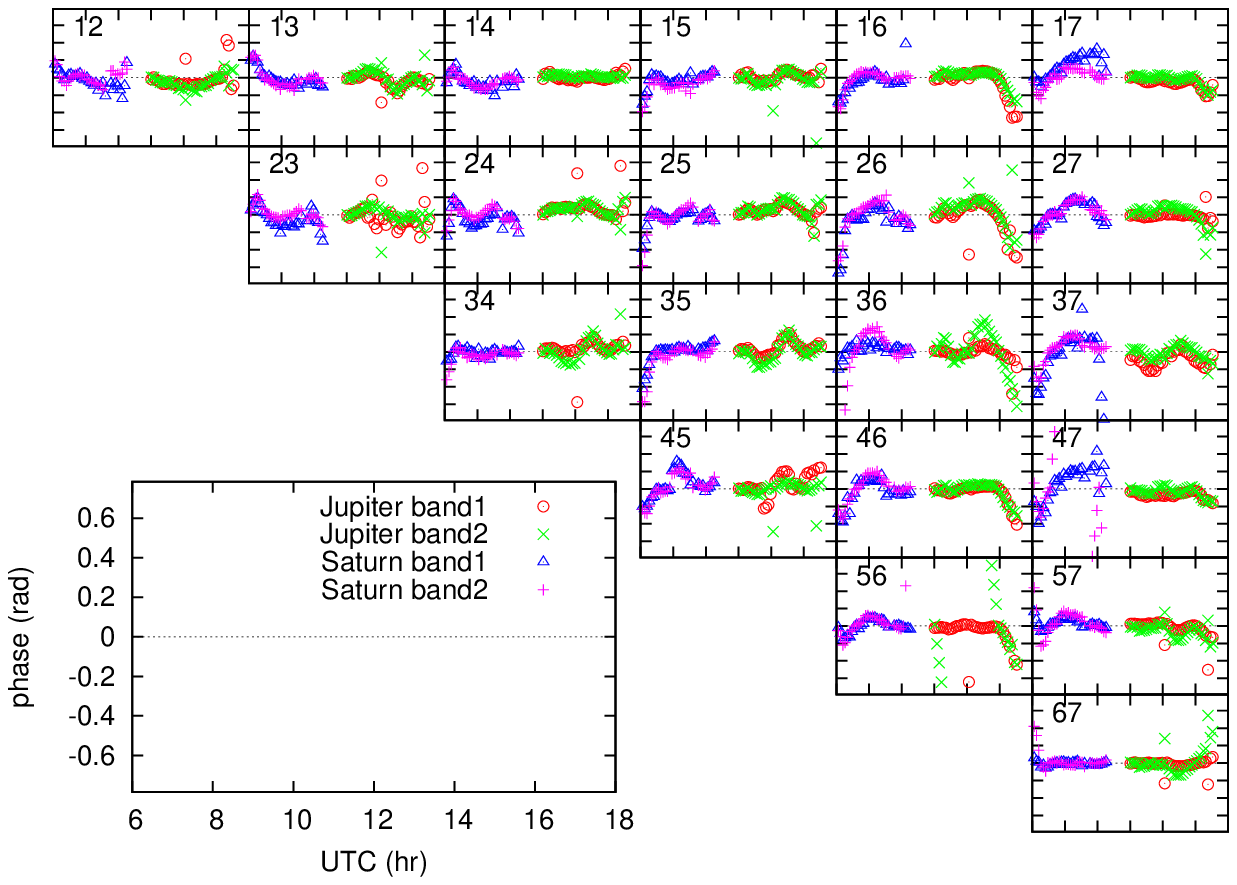}
  \caption{
   Visibilities recovered from a repeated two-patch tracking of Saturn (open 
   triangles and plus symbols) and Jupiter (open circles and cross symbols) 
   in the local time range 8~pm to 8~am (UT 6hr to 18hr). The upper panel 
   shows the relative gain fluctuation, and the lower panel shows the phase 
   variation. Both are shown for the XX correlations. The number in the 
   upper-left corner of each subplot indicates the antenna combination. 
   The scale of the plot and the UT time range are indicated at the 
   bottom left subplot of each panel.
  }
  \label{fig:stability}
\end{figure*}

The stability of the system was examined by measuring the variation in
visibilities for a few bright planets during local times 8~pm to 8~am,
as normally used for observing. 
For this test, the ephemerides of the planets were taken at the
beginning of each track but not updated during the observation. This
causes a pointing error that increases to about an arcminute over
12~hours. 
To account for this, two sets of visibility data for each
planet were chosen as calibrating events. A linear interpolation was
used to remove the linear drift. For data without bracketing events,
the nearest calibration was used. 
Fig.~\ref{fig:stability} shows an example of a stability measurement.
The gain stability was found to have an RMS variation around 5\%, and
the phase to have an RMS variation around 0.1~rad. The measurements also
reveal that the phase response is more sensitive to changes in environment
than the gain response, especially in the first hour after shelter
opening. 

\begin{figure}
  \plotone{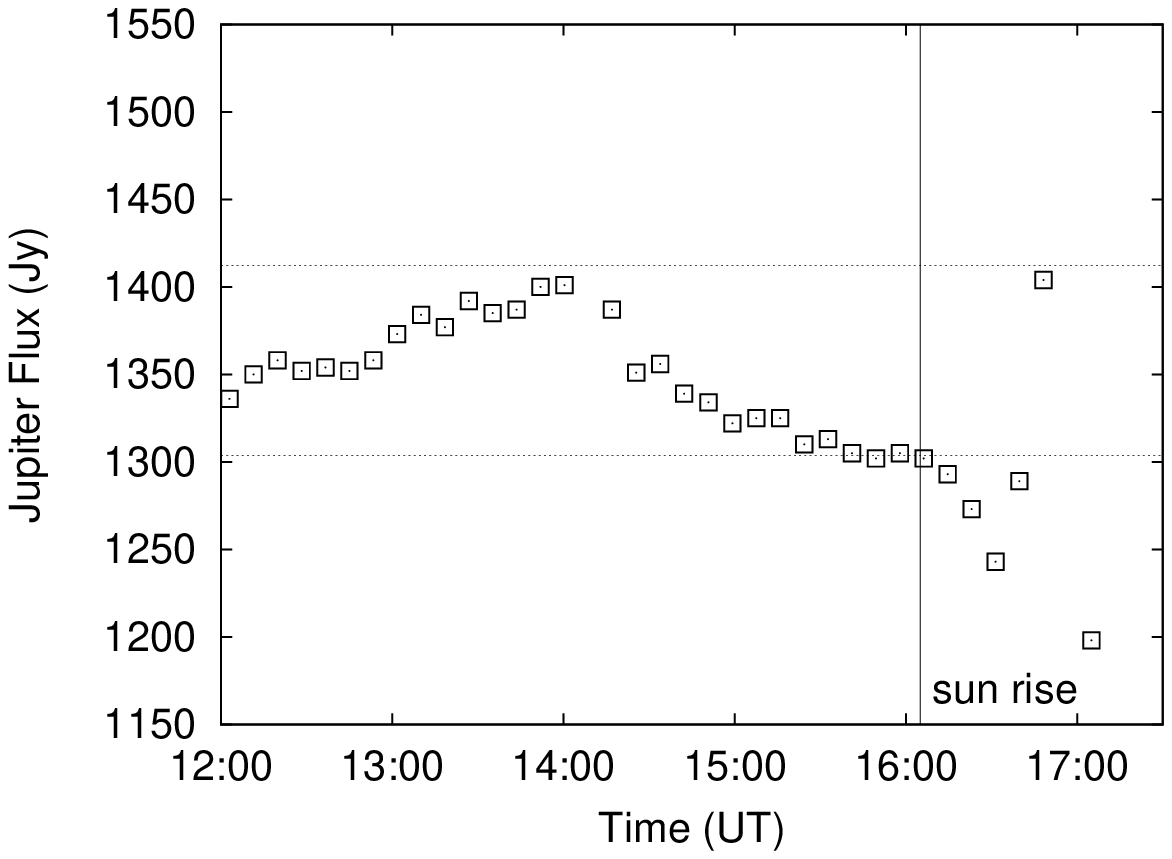}
  \caption{Jupiter flux recovered from the stability measurements in
    Fig.\ref{fig:stability}. Two horizontal dotted lines 
    indicate $\pm 4$\% of the expected Jupiter flux.
  }
  \label{fig:jupflux}
\end{figure}

Fig.\ref{fig:jupflux} plots the flux of Jupiter recovered from the data 
set used in Fig.\ref{fig:stability}. Data was calibrated by the first
measurement at UT~12h (not plotted). The recovered flux varied within
$\pm 4$\% of the  expected flux until sunrise. Calculation of
the calibrator flux is discussed in \S\ref{sec:calibrator}.

Based on the stability measurements, we chose to use a calibration interval of 
two to three hours, to give calibrations good to about 5\% in gain and 0.1~rad 
in phase for each baseline. Calibration requires $\sim 10$\% of telescope 
observing time.

\subsection{Minimizing Instrumental and Ground Pickup\label{sec:offset}}
The signal in the lag output should be constant when AMiBA tracks a
source.  However, the weak signal we measure is susceptible to
slowly-varing contamination. 
The system is designed with a phase switching and demodulation scheme to 
remove contamination such as a common mode leakage in the IF paths.
To modulate the signal, we use a PIN switch to change the LO between two 
carefully adjusted delay lines, and thus changes the phase of IF signal by 
180$^\circ$. The demodulation is done in the readout process. The aim 
was to remove contamination between the down-conversion mixer and the
correlator readout.

However, mixers in the correlator can pick up higher-order signals such as 
$|E_1|^2|E_2|^2$ in addition to their nominal output, which is proportional to 
$E_1 E_2^\ast$ or $|E|^2$, where $E_i$ stands for the voltage from Ant$_i$. 
If the power of the IF signal is modulated by the phase switching pattern, then
this higher order response can generate an output that is coherent with the 
demodulation pattern and becomes a spurious signal. 
This effect is, indeed, seen in AMiBA, where phase switching of the
LO can result  
in a power difference as large as $0.3$~dB. The LO power modulation is 
carried through to the IF in varying amounts depending on the mean LO power 
level at the mixer. The IF modulation can be undetectable for optimally-tuned 
mixers, but is up to $3$~dB for under-pumped mixers.

To reduce the IF modulation, the LO drive level is optimized for minimum 
conversion loss for each mixer. Modulation of the LO power would then have the 
least impact on the IF power. 
Furthermore, to reduce the drifting of LO level with ambient temperature 
changes, the final amplifier and frequency doubler in the LO chain are operated
in the soft saturation regime. Additional protection is provided by 
temperature controls installed before the 2009 observing season.

Spurious signals external to the system, such as ground pickup, will
still affect the data. We used a subtraction scheme similar to the one
used by CBI \citep{pad02} to suppress the slowly-varying signals.  
In practice, we have found that the spurious signal in individual
patches in $\sim$5~hrs integration can be as high as
$\pm 7$~Jy/beam, but that after subtraction a cluster with brightness 0.3~Jy 
can be detected at $9\sigma$ level in 11~hours (i.e., with 5.5~hours on-source 
integration). The observing strategy and the data analysis are given in 
\citet{wu08}.

\section{ACHIEVED SYSTEM PERFORMANCE\label{SEC:PERFORMANCE}}
\subsection{Overall Efficiency\label{sec:efficiency}}
For each baseline, losses include the antenna loss, antenna misalignment, 
and the correlation loss. The antenna loss is mainly related to our Cassegrain 
design. \citet{koc08c} calculate the overall antenna efficiency to be 0.58 
for the 1.2~m and 60~cm dishes alike. The loss originates mainly from three 
factors: the illumination efficiency, the secondary blockage, and the forward
spillover. The AMiBA feeds provide a Gaussian illumination pattern and
the 
the reflectors are designed to have a -10.5~dB edge taper. 
Compared to an uniformly illuminated reflector, only 90\% of the
dish is effectively used. The shadow of the secondary mirror blocks about 8\% 
of the collecting area, giving a loss of 0.92. The edge of the secondary
corresponds to the edge of the primary mirror. Therefore the illumination is 
either reflected by the primary to the sky or is emitted toward the sky 
direcly. The latter part constitutes about 22\% of the energy giving the 
forward spillover factor of 0.78. We favored a design with slightly
worse 
forward spillover but little to none backward spillover (illumination toward
the ground) to reduce system temperature.

The antenna misalignment consists of the
mechanical installation error and the dynamical deformation of the
platform. The former error was measured to be around $3^\prime$ during
the 2007 observing season (Wu et al. 2009, in preparation) and will 
be improved for future
observations. The latter error was inferred from photogrammetry
measurements of the platform surface to be less than 
1\arcmin\ \citep{koc08a}. The two errors together attenuate the
primary beam by 
2\%. Antenna misalignment may also cause pointing errors for some
baselines. This effect is not considered in individual baseline
efficiencies but will be considered in the array efficiency. There is
also a loss of efficiency from the noise contributed by the rejected
correlations in the analog correlator. The estimated correlation efficiency 
from this effect is 0.81. 

When combining baselines from the entire array, pointing error and
system stability also lower the efficiency by degrading
the coherence of signal from different measurements. The pointing
error is less than 0.4\arcmin\ \citep{koc08a} and decreases the
efficiency by less than 2\%. The large alignment error, on the other
hand, contributes as much as 12\% loss in the 2007 and 2008 observations. As
described in \S\ref{sec:stability}, the system stability is
approximately $\pm 5\%$ in gain and $\pm 0.1$~rad in phase. Taken over
all baselines this results in a reduction of signal by about 2\%.  
Table~\ref{tab:efficiency} summarizes the major losses in the system.

\begin{deluxetable}{ll}
  \tablecaption{Summary of Losses of The System\label{tab:efficiency}}
  \tablehead{
    \colhead{Systematics} & \colhead{Efficiency}
  }
  \startdata
  Antenna illumination\tablenotemark{a} & 0.90 \\
  Antenna blockage\tablenotemark{a} & 0.92 \\
  Antenna spillover\tablenotemark{a} & 0.78 \\
  Antenna others effects\tablenotemark{a,b} & 0.90 \\
  \hspace{1cm}Antenna total & $0.90\times0.92\times0.78\times0.90=0.58$ \\
  Alignment\tablenotemark{b} & 0.98 \\
  Correlation\tablenotemark{a} & 0.81 \\
  \hspace{1cm}{\bf Overall Baseline} & {\bf $0.58\times0.98\times0.81=0.46$} \\
  &  \\   
  Deformation/Pointing\tablenotemark{b} & 0.88 \\
  Stability\tablenotemark{b} & 0.98 \\
  \hspace{1cm}{\bf Overall Array} & {\bf $0.46\times0.88\times0.98=0.40$}
  \enddata
  \tablenotetext{a}{The facor is based on theoretical calculation.}
  \tablenotetext{b}{The facor is derived from measured quantities.}
\end{deluxetable}

The baseline efficiency has been checked by comparing the signal-to-noise ratio
(SNR) of Jupiter's fringe to the ratio of Jupiter's antenna temperature and the
system temperature. 
$\eta_{bl} = \mathrm{SNR}_{Jup} / ( \frac{T_{a,Jup}}{T_{sys}} \sqrt{ B_{eff} \frac{t_{rec}}{2} } )$, 
where $T_{a,Jup}$ is the antenna temperature of Jupiter, typically around 0.1K 
for the 60cm dishes, and $\mathrm{SNR}_{Jup}$ is the SNR of Jupiter under the 
corresponding recording time $t_{rec}$ ($=0.452$~sec currently). An average 
effective bandwidth of $B_{eff}\sim 10$~GHz was assumed in the calculation. 
The measured efficiency scatters from 0.2 to 0.5 with an error bar of 
approximately 0.2. The error originates mainly from the noise estimation of 
the signal-dominant fringe, the occasional large readout noise, and also the 
variation of effective bandwidth. The overall array efficiency will be covered
in \S\ref{sec:snr}.

\subsection{Calibrator\label{sec:calibrator}}
The raw visibilities recovered from the lag data have the systematic
losses discussed above and are further affected by instrumental delay,
gain drift, phase variation as well as the imperfect lag-to-visibility
transformation. We calibrate visibilities by interspersed
two-patch observations of a planet. 
This converts our visibility amplitudes to flux density units and references 
the phase to the calibrating planet position.

Taking planet data with the subtraction scheme, and applying the same
calibration scheme used for cluster data (one calibration about every
three hours), we find that the recovered peak flux in the image domain
shows an RMS scatter of about 3\%. 

\begin{figure}
  \plotone{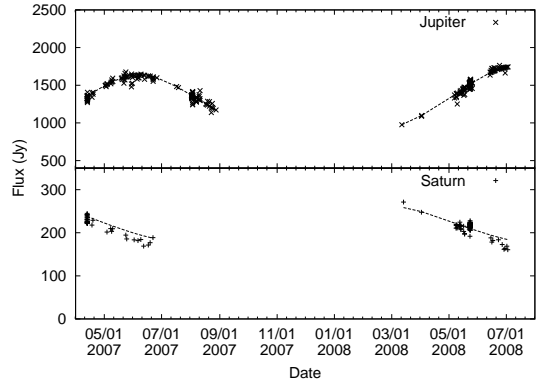}
  \caption{The recovered flux densities of our main calibrators,
    Jupiter (plus symbols)  
    and Saturn (cross symbols) from the observations in 2007 and 2008. Data 
    were calibrated by the first Jupiter measurement in the same
    night, or the 
    nearest Jupiter measurement. The expected flux densities are
    plotted as dashed and  
    solid lines for Jupiter and Saturn respectively. Note that Saturn's ring 
    was not taken into account in the flux density estimation.
  }
  \label{fig:calflux}
\end{figure}

The flux densities of the planets are calculated from published disk
brightness temperatures and the apparent angular sizes assuming a
black-body spectrum. We adopt the values: Jupiter 171.8$\pm$1.7 K
\citep{pag03,gri86}, Saturn 149.3$\pm$4.1 K \citep{uli81}, and Mars
206.8$\pm$5.6 K \citep{uli81}. 
Fig.\ref{fig:calflux} shows the recovered flux of the main calibrators (Jupiter
and Saturn) in the 2007 and 2008 observations and the expected flux from 
calculation. For the phase reference and flux standard we use the
first Jupiter measurement on each night when Jupiter is observed, as this
provides the best  
calibration for the earlier Saturn measurements. The scatter of the Jupiter 
flux densities agrees with the scatter in one night as shown in
Fig.\ref{fig:jupflux}.  
The flux density of Saturn is systematically lower than the calculated
value by  
approximately $5\%$. It was verified that the lower flux was not the
result of  
an error in flux standard. This was checked by artificially setting
the phase 
error to zero in the calibrated visibilities of Saturn and forming an image. 
The recovered flux density displays no systematic offset from the
calculated level, to within the error bound. 
The apparent deficit in the flux density for Saturn is likely due to 
insufficient phase calibration in the 2007 data, when calibrations
more than two~hours apart were often used.
Note that the effect of Saturn's ring is not included 
in the calculation of Saturn's flux density. The ring inclination was
about -15~deg for the 2007 observations and about -10~deg for the 2008
observations.  
We estimate that our flux density scale is good to about $\pm 5$\% in absolute 
terms. Based on Jupiter's flux scale, our measurement of Saturn's disk 
brightness temperature is $149^{+5}_{-12}$~K.

A final note about calibration is on the difference of spectra between the
calibrator and the SZ effect. Our primary calibrators, Jupiter and Saturn, 
are dominated by thermal emission near 94~GHz. The black-body spectrum favors
slightly higher frequency than does the SZE spectrum. Regardless of the choice
of assumed passband shape, the difference of effective central frequency, 
defined by 
$f_c = [\int_{{\mathrm band}} R(f)S(f)fdf]/[\int_{{\mathrm
band}}R(f)S(f)df]$ with $S(f)$ being the source spectrum and $R(f)$
being the passband, is less than 1.5\% of the bandwidth. Its effect on
the calibration is thus negligible compared to other errors.

\subsection{Noise Integration\label{sec:snr}}
\begin{figure*}
  \centering
  \includegraphics[totalheight=5.5in]{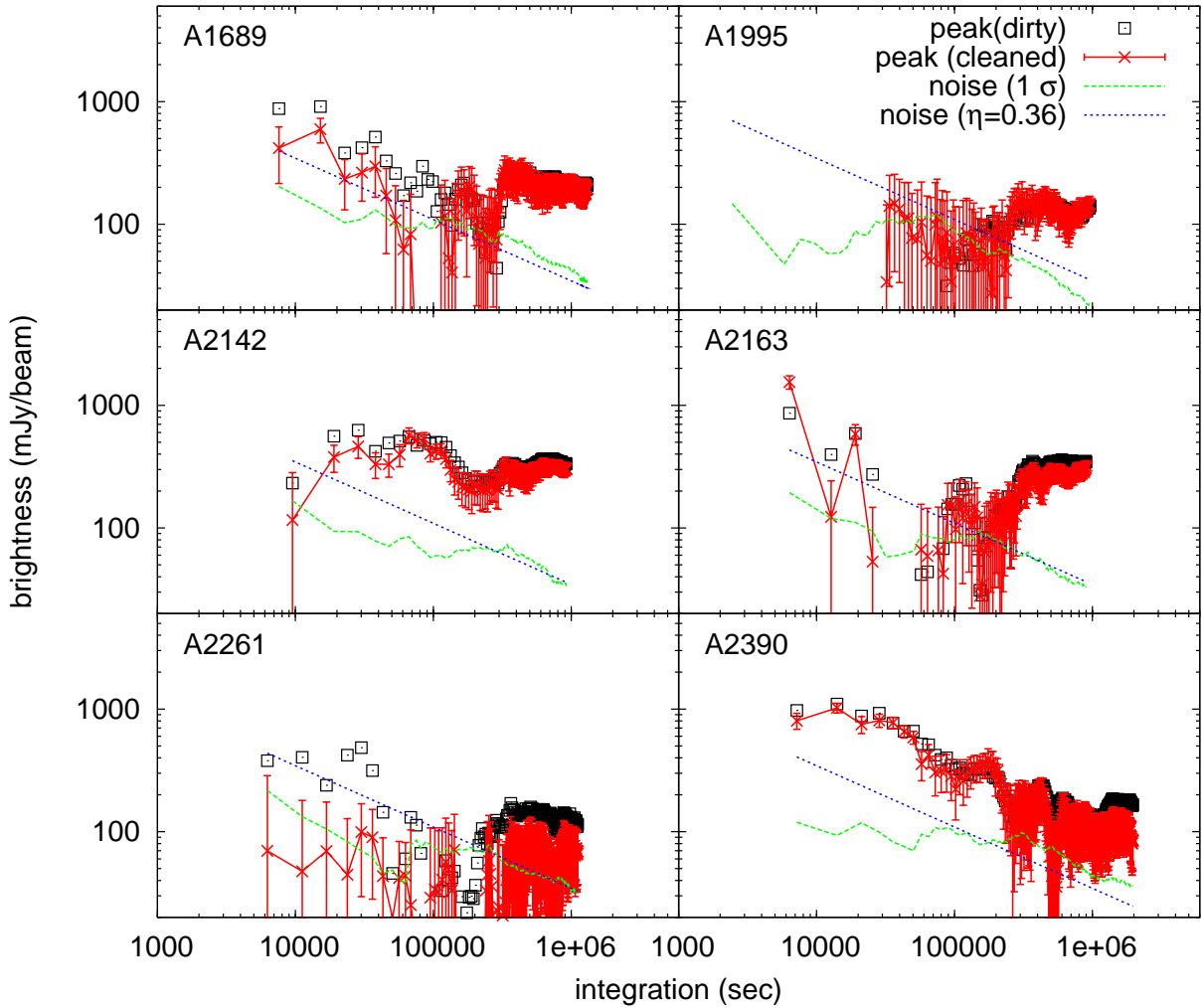}
  \caption{
    The signal and noise plotted against effective integration time
    for our sample. The black open squares and the red solid line with 
    error bars represent the signal measured in dirty and cleaned maps 
    respectively and have been multiplied by -1 in the plot. The
    green dashed line shows a noise estimate from the cleaned maps 
    (see \S\ref{sec:snr}). 
    The blue dotted line shows the expected noise level given
    the effective integration time, an overall system efficiency of 0.36,
    and a system temperature of 90~K.
    $10^6$~sec of effective integration on the horizontal axis corresponds
    to roughly 3.3~hours on-source obervation time.
    The final signal-to-noise ratios (SNR) in the cleaned images are 
    6.0 for A1689, 6.4 for A1995, 13.7 for A2142, 5.2 for A2163, and
    6.6 for A2390).
   }
  \label{fig:snr_int}
\end{figure*}

Based on the signal-to-noise ratio (SNR) of Jupiter's fringe and the
discussion in \S\ref{sec:efficiency}, we find that the array has an
overall efficiency of about 0.4. The improvement in sensitivity with 
integration time depends critically on the removal of
spurious signals using the subtraction scheme and subsequent data
flagging. To verify the sensitivity, we examine the variation of
signal and noise in the reconstructed map with integration time in
Figure \ref{fig:snr_int}. The integration time here refers to the
accumulation of observing time spent in each individual visibility
band. For example, when the telescope tracks a source for 3
minutes, the total integration time for 21 baselines, 2 polarizations,
and 2 bands is
$t_{tot}=180$~sec~$\times21\times2\times2=15120$~sec. Since the
visibilities are used with non-uniform weights in forming the image,
we calculate an effective integration time, which for 3 minutes
on-source integration is defined as $t_{eff}=\frac{(\sum_i
w_i)^2}{\sum_i {w_i}^2}\times180$~sec, where $w_i$ denotes the weighting
given to each data set. In this analysis, a natural weighting is
adopted. 

Since the noise comes from two patches and the signal comes from only one, the 
point source sensitivity can be estimated by 
$\sigma = \frac{2kT_{sys}}{\eta_{all}A_{phys}}\frac{1}{\sqrt{t_{eff}B_{ch}}}$, 
where $T_{sys}=100$~K, $B_{ch}=5$~GHz, $\eta_{all}$ is the overall efficiency, 
and $A_{phys}$ is the physical collecting area.

The signals are read from the source position in the reconstructed
dirty images with different integration times, with the source position 
determined from the final image. A CLEAN \citep{hog74} procedure is
applied to the inner 21.6\arcmin~box, which roughly corresponds to
the FWHM of the primary beam. The cleaned signal at the source position
is also recorded, while the residual noise is measured in the 1 deg
image excluding the inner clean region. 

Figure~\ref{fig:snr_int} shows that an efficiency $\eta=(0.36\pm0.04)$
is more  representative for the current data, giving a point source
sensitivity of $63 \pm 7$~mJy in 1~hour of on-source integration 
when the subtraction scheme is applied. 
Abell~2390 has a higher noise, and we believe this may be caused by the 
presence of point sources in the fov. \citet{liu08} investigate the 
contamination by point sources and the primary CMB.

\section{Conclusion\label{SEC:CONCLUSION}}
To detect galaxy clusters with the AMiBA, we must achieve system stability on 
timescales of hours.  We have optimized the performance of AMiBA by measuring 
and compensating for the delays between antennas, and using beam switching 
techniques to cancel out instrumental and environmental effects.  Planet 
calibrations provided corrections for passband response.  Overall efficiency 
for AMiBA was $\eta_{all}=0.36\pm0.04$, with a major loss from the antenna 
efficiency, $\eta_{ant}=0.58$.

Using a system temperature of $90$~K, an effective bandwidth of $5$~GHz per 
band, and an overall efficiency of $0.36\pm0.04$, the point source sensitivity 
of AMiBA in 1~hour of on-source integration ($t_{eff}=302400$~sec) is
found to be about 
$63 \pm 7$~mJy when the subtraction scheme is applied. The effective
integration 
is about 60\% of the on-source integration time. The loss of 40\% of observing 
time is mostly due to lower weighting applied to some receivers or baselines 
experiencing hardware problems. There were very few observations made when the 
weather was not good in the 2007 and 2008 observing season.

The flux density error consists of the calibrator flux scale
uncertainty of $\pm 5$~\%,  
and the cross-calibration error $\pm 3\%$, which also includes the 
lag-to-visibility flux scale error of $\pm 2\%$. The latter two errors
are well
below the thermal noise in all clusters observed during
2007 and 2008. Investigation of noise in cleaned images shows that longer
integration, aimed at measuring primordial CMB fluctuation, will be
promising. 

The resulting successful detections of clusters have led to a number of
scientific results including a measurement of the Hubble constant and the 
study of the hot gas distribution in the clusters.  These are discussed further
in the companion papers.

\acknowledgments
{\bf Acknowledgments}  
Capital and operational funding for AMiBA came from the MoE and the NSC as part
of the Cosmology and Particle Astrophysics (CosPA) initiative. Additional 
funding also came in the form of an Academia Sinica Key Project. Support from 
the STFC for MB is also acknowledged.



\begin{thebibliography}{}
\bibitem[Bond et al.(2005)]{bon05} Bond, J. R. et al. 2005, ApJ, 626, 12
\bibitem[Chen et al.(2009)]{mtc08} Chen, M.-T. et al. 2009, ApJ, submitted
\bibitem[Goldstein et al.(2003)]{gol03} Goldstein, J. H. et al. 2003, ApJ, 599, 773
\bibitem[Griffin et al.(1986)]{gri86} Griffin, M. J. et al. 1986, Icarus, 65, 244
\bibitem[Ho et al.(2009)]{ho08} Ho, P. T. P et al. 2009, ApJ, accepted (astro-ph/0810.1871)
\bibitem[Hogbom (1974)]{hog74} Hogbom, J. A. 1974, A\&AS, 15, 417
\bibitem[Huang et al.(2009)]{hcw08} Huang, C.-W. L. 2009, ApJ, submitted
\bibitem[Huang et al.(2008)]{ydh08} Huang, Y.-D. et al. 2008, SPIE, 7012, 70122H
\bibitem[Koch et al.(2009a)]{koc08a} Koch, P. M. et al. 2009a, ApJ accepted
\bibitem[Koch et al.(2009b)]{koc08b} Koch, P. M. et al. 2009b, ApJ, submitted
\bibitem[Koch et al.(2009c)]{koc08c} Koch, P. M. et al. 2009c, in preparation
\bibitem[Kuo et al.(2004)]{kuo04} Kuo, C.-L. et al. 2004, ApJ, 600, 32
\bibitem[Li et al.(2004)]{ctl04} Li, C.-T. et al. 2004, SPIE, 5498, 455
\bibitem[Lin et al.(2004)]{kyl04} Lin, K.-Y., Woo, Tak-Pong, Tseng, Yao-Huan, Lin, Lihwai, and Chiueh, Tzihong 2004, ApJ, 608, 1L
\bibitem[Lin et al.(2009)]{kyl08} Lin, K.-Y. et al. 2008, SPIE, 7012, 701207
\bibitem[Liu et al.(2009)]{liu08} Liu, G.-C. et al. 2009, ApJ, submitted
\bibitem[Nishioka et al.(2009)]{hn08} Nishioka, H. et al. 2009, ApJ accepted (astro-ph/0811.1675)
\bibitem[Padin et al.(2002)]{pad02} Padin, S. et al. 2002, PASP, 114, 83
\bibitem[Page et al.(2003)]{pag03} Page, L. et al. 2003, ApJS, 148, 39
\bibitem[Pearson et al.(2003)]{pea03}Pearson, T. J. et al. 2003, ApJ, 591, 556
\bibitem[Umetsu et al.(2009)]{ku08} Umetsu, K. et al. 2009, ApJ accepted (astro-ph/0810.0969)
\bibitem[Wu et al.(2009)]{wu08} Wu, J.-H. P. et al. 2009, ApJ accepted (astro-ph/0810.1015)
\bibitem[Ulich (1981)]{uli81} Ulich, B. L. 1981, AJ, 86, 1619
\end{thebibliography}
\end{document}